\documentclass[aps,prb,twocolumn,showpacs]{revtex4}
\usepackage{amssymb}

\usepackage{graphicx}

\begin{document}

\title{Strong coupling superconductivity and prominent superconducting fluctuations in the new superconductor Bi$_4$O$_4$S$_3$}

\author{LI Sheng$^1$, YANG Huan$^{1,\dag}$, FANG DeLong$^1$, WANG ZhenYu$^2$, TAO Jian$^1$, DING XiaXin$^1$, and WEN HaiHu$^{1,*}$}

\affiliation{$^1$Center for Superconducting Physics and Materials, National Laboratory of Solid State Microstructures and Department of Physics, Nanjing University, Nanjing 210093, China}
\affiliation{$^2$National Laboratory for Superconductivity, Institute of Physics and National Laboratory for Condensed Matter Physics, Chinese Academy of Sciences, Beijing 100190, China}

\begin{abstract}
Electric transport and scanning tunneling spectrum (STS) have been
investigated on polycrystalline samples of the new superconductor
Bi$_4$O$_4$S$_3$. A weak insulating behavior in the resistive
curve has been induced in the normal state when the
superconductivity is suppressed by applying a magnetic field.
Interestingly, a kink appears on the temperature dependence of
resistivity near 4$\;$K at all high magnetic fields above 1$\;$T
when the bulk superconductivity is completely suppressed. This
kink associated with the upper critical field as well as the wide
range of excess conductance at low field and high temperature are
explained as the possible evidence of strong superconducting
fluctuation. From the tunneling spectra, a superconducting gap of
about 3$\;$meV is frequently observed yielding a ratio of
$2\Delta/ k_B T_\mathrm{c} \approx 16.6$. This value is much
larger than the one  predicted by the BCS theory in the weak
coupling regime ($2\Delta /k_\mathrm{B}T_\mathrm{c} \approx
3.53$), which suggests the strong coupling superconductivity in
the present system. Furthermore, the gapped feature persists on
the spectra until 14$\;$K in the STS measurement, which suggests a
prominent fluctuation region of superconductivity. Such
superconducting fluctuation can survive at very high magnetic
fields, which are far beyond the critical fields for bulk
superconductivity as inferred both from electric transport and
tunneling measurements.\end{abstract}


\maketitle

Superconductivity is induced by quantum condensation of a large
number of paired electrons. The pairing is supposed to be
established between the two electrons with opposite momentum and
spins by exchanging the phonons. According to the
Bardeen-Cooper-Schrieffer (BCS) theory, a linear relationship
between the electron pairing gap $\Delta$ and superconducting (SC)
critical temperature $T_\mathrm{c}$, i.e.,
$2\Delta/k_\mathrm{B}T_\mathrm{c}=3.53$ exists in the weak
coupling regime. In recent years, the original proposal about the
pairing based on the electron-phonon coupling has been gradually
replaced by the more exotic pairing mechanism, such as through
exchanging the magnetic spin fluctuations, and $T_\mathrm{c}$ can
be increased to a higher level. The SC pairing mechanism of the
cuprates \cite{1} and the iron pnictides \cite{2}, although not yet being
settled down completely, should have a close relationship with
strong correlation effect. In cuprates \cite{3} and pnictides \cite{4}, the
ratio $2\Delta/k_\mathrm{B}T_\mathrm{c}$ is generally much larger
than that predicted by the weak coupling BCS theory, implying the
strong coupling superconductivity (in some conventional BCS
superconductors, like Pb, Hg and PbBi, the coupling is also strong
\cite{5}). Another important issue for cuprates is that the electron
correlation in the normal state may induce strong superconducting
fluctuation which has been widely investigated \cite{6,7,8,9,10}. However,
the fluctuation region in iron pnictides seems not that wide
\cite{11,12}. SC fluctuation was used to be associated with another
characteristic property, the so-called pseudogap effect in
cuprates \cite{13}. However the SC fluctuation temperature was proved
to be different from the pseudogap temperature by further
experiments \cite{14,15}. Although pseudogap has a very close
relationship with superconductivity, there is however no consensus
yet on the origin of it in cuprates. The scanning tunneling
spectra show a continuous evolution from the SC gap feature to the
pseudogap feature which exists at temperatures far above
$T_\mathrm{c}$ \cite{16}. High-$T_\mathrm{c}$ superconductivity should
have a close relationship with the correlation effect of electrons
from the 3d orbitals \cite{17,18,19}. In the heavy Fermion systems \cite{20}
and organic materials \cite{21}, a similar conclusion may be achieved.
In this regard, it is natural to explore new superconductors with
possible higher transition temperatures in compounds with
transition metal elements in which the electron correlations could
be strong. The electrons in the p-orbital, especially those from
the 5p or 6p orbits, are normally assumed to have weak repulsive
potential and quite wide band width, and hence have very weak
correlation effect. It would be surprising to find out
superconductivity with exotic feature in the p-orbital based
compounds.

Recently, Mizuguchi et al. \cite{22} discovered superconductivity with
$T_\mathrm{c}^\mathrm{onset} = 8.6\;$K (determined from the point
where resistivity deviates from the linear extrapolation of
normal-state value) and $T_\mathrm{c}^\mathrm{zero} = 4.5\;$K in
the so-called BiS$_2$ based compound Bi$_4$O$_4$S$_3$. This
compound has a layered structure with the space group of I4/mmm or
I-42m. Shortly the same group reported superconductivity at about
10.6 K in another system LaO$_{1-x}$F$_x$BiS$_2$ by doping
electrons into the material through substituting oxygen with
fluorine \cite{23}. By replacing the La with Nd or Ce, other new
BiS$_2$ based materials were reported \cite{24,25}. Quickly followed is
the theoretical work based on the first principles band structure
calculations \cite{26}, which predicts that the dominating bands for
the electron conduction as well as for the superconductivity are
derived from the Bi 6$p_x$ and 6$p_y$ orbits. In this paper, we
present a set of data for an intensive study on the transport and
scanning tunnelling spectroscopy (STS) measurements of the new
superconductor Bi$_4$O$_4$S$_3$. Our results clearly illustrate
the strong coupling superconductivity and prominent
superconducting fluctuation in this interesting superconductor.

\section{Experiments}

The polycrystalline samples were synthesized by using a two-step
solid state reaction method. Firstly, the starting materials
bismuth powder (purity 99.5$\%$, Alfa Aesar) and sulfur powder
(purity 99.99$\%$, Alfa Aesar) were mixed in a ratio of 2:3,
ground and pressed into a pellet shape. Then it was sealed in an
evacuated quartz tube and followed by annealing at
$500\,^{\circ}$C for 10 hours. The resultant pellet was smashed
and ground together with the Bi$_2$O$_3$ powder (purity 99.5$\%$,
Alfa Aesar) and sulfur powder, in stoichiometry as the formula
Bi$_4$O$_4$S$_3$. Again it was pressed into a pellet and sealed in
an evacuated quartz tube and burned at $510\,^{\circ}$C for 10
hours. Then it was cooled down slowly to room temperature. The
second step was repeated in achieving good homogeneity. The
resultant sample looks black and very hard. We cut the sample and
obtained a specimen with a rectangular shape for the resistive
measurements. The resistivity was measured with Quantum Design
instrument PPMS-16T. The temperature stabilization was better than
0.1$\%$ and the resolution of the voltmeter was better than
10$\;$nV. The magnetization was detected by the Quantum Design
instrument SQUID-VSM with a resolution of about $5 \times
10^{-8}\;$emu. The sample was shaped as a bar with a typical size
of $2 mm\times2 mm\times0.5 mm\;$ for the STS measurements. Since
the sample is very hard, which allows us to polish the sample
surface and obtain a shiny and mirror-like surface. The top
surface was polished by sandpapers with different grit sizes
(smallest of ISO P10000). The tunnelling spectra were measured
with an ultra-high vacuum, low temperature and high magnetic field
scanning probe microscope USM-1300 (Unisoku Co., Ltd.). In STS
measurements, Pt/Ir tips were used. The set points of the bias
voltage and tunneling current are 40$\;$mV and 100$\;$pA
respectively to fix the tip height in topographic mode. Then the
differential conductivity was recorded while the bias voltage was
swept with the tip held at a fixed vertical distance with
$z\;$-piezo-feedback off for the STS measurements. There is no
atomically resolved topography measured on the sample since it is
a polycrystalline one. The roughness of the surface for STS
measurements is about 2$\;$nm, while on some flat surface of a
grain the roughness is about 0.5$\;$nm locally. The STS spectra
are repeatable at different positions in one grain. In reducing
noise of the differential conductance spectra, a lock-in technique
with an ac modulation of $0.1\;$mV at $987.5\;$Hz was typically
used.
\section{Results}

\begin{figure}
\includegraphics[width=8cm]{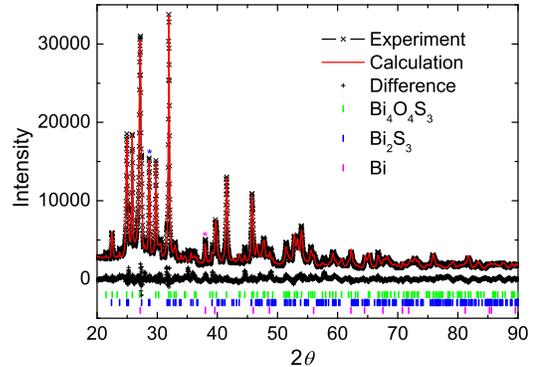}
\caption {(color online)  The XRD pattern (symbols) of
Bi$_4$O$_4$S$_3$ refined (red solid lines) by Topas (Bruker-D8).
It is clear that the main phase is Bi$_4$O$_4$S$_3$ containing
Bi$_2$S$_3$ impurity marked by a blue star and Bi by a red star.
The ratio among the three phases Bi$_4$O$_4$S$_3$:Bi$_2$S$_3$:Bi
is about 80:15:5.} \label{fig1}
\end{figure}

The crystallinity of the sample was checked by x-ray diffraction
(XRD) with the Brook Advanced D8 diffractometer with Cu K$\alpha$
radiation. The analysis of XRD data was done by the softwares
Powder-X and Topas. The XRD pattern looks very similar to that
reported by Mizuguchi et al. \cite{22}. The result of Rietveld fitting
was done with the Topas program in Fig.~\ref{fig1}, yielding a
80$\%$ volume of Bi$_4$O$_4$S$_3$ with 20$\%$ of impurities which
are mainly Bi$_2$S$_3$(15$\%$) and Bi(5$\%$).

\begin{figure}
\includegraphics[width=8cm]{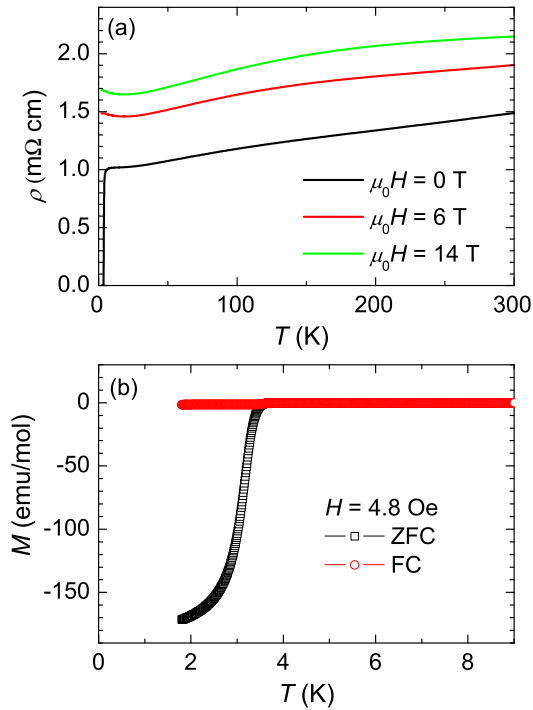}
\caption {(color online) (a) Temperature dependence of resistivity
measured at three magnetic fields: $\mu_0H = $0, 6 and 14 T. It is
clear that, beside a moderate magnetoresistance effect, a weak
insulating behavior is induced by the magnetic field. (b) The
temperature dependence of magnetization near the SC transition
measured with $H = 4.8\;$Oe in the field-cooled (FC) and
zero-field-cooled processes (ZFC).}\label{fig2}
\end{figure}

\begin{figure}
\includegraphics[width=8cm]{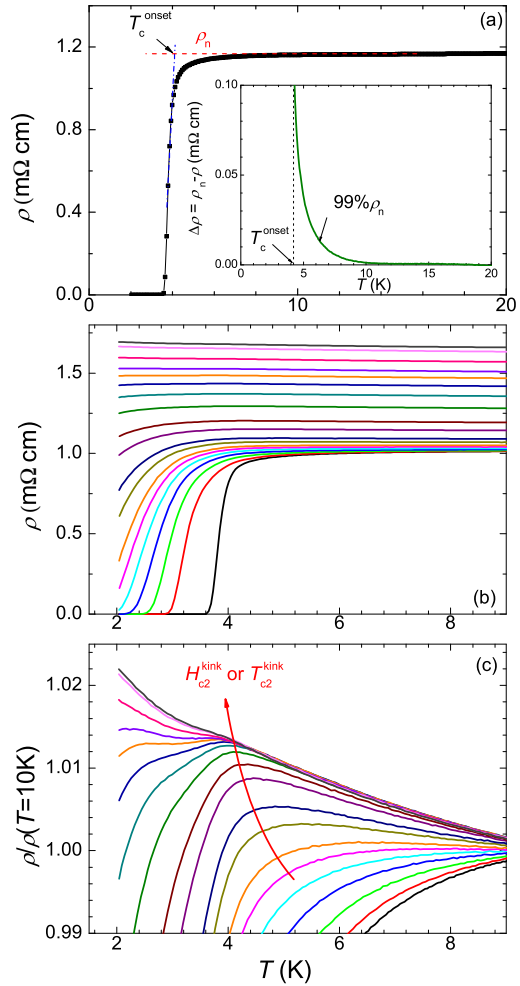}
\caption {(color online) (a) Temperature dependence of resistivity
in Bi$_4$O$_4$S$_3$ at zero field near the resistive transition.
The onset transition temperature $T_\mathrm{c}^\mathrm{onset}$ at
the crossing point of the normal state background (guided by a red
dashed line) and the extrapolated line of the steep resistive
transition part (guided by a blue dash-dot line) is $4.2\;$K. The
resistance difference between the resistive curve and the normal
state background extends to a very high temperature shown in the
inset, which suggests that the SC fluctuation may be strong in
this material. (b) An enlarged view for the temperature dependence
of resistivity at magnetic fields of (from bottom to top) 0, 0.1
to 0.6 T with increments of 0.1 T; 0.8, 1, 1.5, 2 to 7 with
increments of 1 T; and 9, 12, 14 T. It is found that the bulk
superconductivity can be quickly suppressed by the magnetic field,
while the onset transition temperature changes slightly,
indicating a strong fluctuation effect. (c) Temperature dependence
of the resistivity (as shown in (b)) normalized at 10 K. A kink
can be clearly seen at about 4 K when the magnetic field is high
and the bulk superconductivity is suppressed completely. The red
arrowed line traces out the evolution from the SC onset transition
in the low field region to a kink at high magnetic
fields.}\label{fig3}
\end{figure}

In Fig.~\ref{fig2}(a) we present the temperature dependence of
resistivity measured at three magnetic fields: $\mu_0H = $0, 6 and
14 T. In addition to the moderate magnetoresistance, one can see
that a weak insulating behavior is induced by the magnetic field.
This weak semiconducting behavior is of course anti-intuitive for
a normal state with Fermi liquid characteristic. A simple
explanation would be that the insulating feature is given by an
adjacent competing order here, once the superconductivity is
suppressed, the latter is getting promoted. However, we should
mention that the insulating behavior starts actually at 25 K (at 6
and 14 T) which is far beyond the SC transition temperature here.
Someone may argue the minimum in the resistivity is caused by some
impurities of Bi$_2$S$_3$, which shows a minima around 25 K \cite{27},
but this may be excluded because we do not see this phenomenon in
zero magnetic field. Another possibility is that the conduction
band has a very shallow band edge, as illustrated by the band
structure calculations \cite{26}. When a magnetic field is applied, the
density of states (DOS) of the spin-up and spin-down electrons
will become asymmetric given by the Zeeman effect. Therefore we
have some polarized electrons which induce the weak insulating
behavior. Clearly this insulating behavior needs to be further
checked, better with single crystals in the future, and to be
explained satisfactorily. In Fig.~\ref{fig2}(b) we present the
magnetization data measured in the zero-field-cooled (ZFC) and the
field-cooled mode (FC). The SC magnetic transition starts at about
3.6 K. Superconducting transition temperature in this paper is a
little lower than that in previous report by Mizuguchi et al.
\cite{22}, probably due to mutual doping between the O and S elements.
From the resistive curve in the transition region as shown in
Fig.~\ref{fig3}(a), the critical temperature taken from $5\%$ of
the normal state resistivity ($T_\mathrm{c}^\mathrm{zero}$ or
$T_\mathrm{irr}$) is 3.7$\;$K. The onset transition temperature
$T_\mathrm{c}^\mathrm{onset}$ determined from the crossing point
of the normal state line and the extrapolation of the steep
transition line is about 4.2$\;$K, while
$T_\mathrm{c}(99\%\rho_\mathrm{n})$ taken from the $99\%$ of the
normal resistance is about 6.2$\;$K. It should be noted that the
excess conductivity region in which the resistivity starts to
deviate from the normal state line (inset of Fig.~\ref{fig3}(a))
can extend to the temperature above 10$\;$K. This excess
conductivity is usually regarded as the SC fluctuation from the
residual Cooper pairs above bulk
$T_\mathrm{c}(99\%\rho_\mathrm{n})$. This superconducting
fluctuation is actually expected by the band structure
calculations which foresee a low dimensionality of the electronic
structure, and can be corroborated by the quickly broadened
resistive transition under magnetic fields, as shown in
Fig.~\ref{fig3}(b). One can see that the transition temperature
with zero resistivity can be suppressed blow 2 K by a magnetic
field as low as 0.4 T, while the bulk superconductivity is
suppressed completely by a magnetic field of 5 T. However, even if
the bulk superconductivity is easily suppressed, a kink appears on
the $\rho$ vs $T$ curve at a high magnetic field where one cannot
see the diamagnetization. If following the onset transition of the
resistivity, as shown in Fig.~\ref{fig3}(c), we can see that the
kink has very close relationship with the upper critical field
$\mu_0H_\mathrm{c2}$ in the low field and high temperature region.
Because it is really difficult to define the temperature below
which $\rho_\mathrm{n}$ first deviates from its high temperature
behavior, we use these kink positions to define the upper critical
fields above 1 T. Surprisingly, this kink stays at about 4 K even
with a magnetic field of 14 T. We interpret this kink as the
temperature below which the residual Cooper pairs exist in the
system even the bulk superconductivity is completely suppressed.
Following the tendency of this kink, a very high critical field
can be expected in the zero temperature limit, which certainly
exceeds the Pauli limit given by $\mu_{0}H_\mathrm{p} =
1.84T_\mathrm{c}$ \cite{28}.

\begin{figure}
\includegraphics[width=8cm]{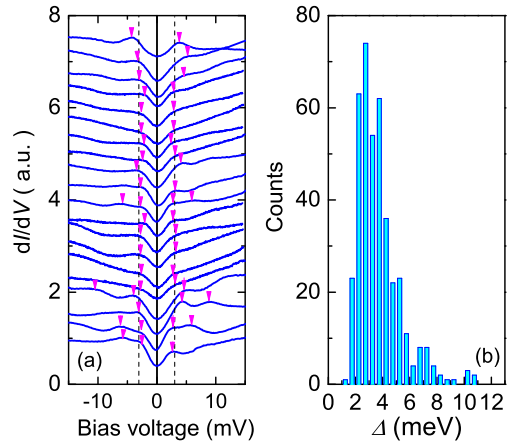}
\caption {(color online) (a) Typical tunnelling spectra measured
at different locations on the polycrystalline samples. The gap
values judged from the peaks or humps on the spectra are marked by
the arrows. One can see that the mean gap value is around 3$\;$meV
as guided by the vertical dashed lines. Some spectra show very
large gap values and even have two-gap features. (b) Statistics on
the local SC gap sizes $\Delta$ of 400 spectra. The gap size
follows the Gaussian distribution with the mean gap value
$\overline{\Delta}=3\;$meV. Such a large gap value suggests the
unconventional superconductivity in Bi$_4$O$_4$S$_3$ with
$T_\mathrm{c}^{\mathrm{onset}}=4.2\;$K. The largest SC gap can
reach a value of about 10 meV} \label{fig4}
\end{figure}

To make further analysis on the superconducting property, we
measured STS spectra on this sample. Several typical STS curves
measured at 1.6$\;$K below the bulk SC transition are shown in
Fig.~\ref{fig4}(a). Most of the spectra are symmetric with very
clear suppression of DOS within a certain energy scale, and clear
coherence peaks can be found on some spectra. However the
coherence peaks on most of the curves are somewhat broad and the
zero-bias conductance values are remarkably large, which maybe due
to the contamination of the surface on this polycrystalline
sample. The gap values determined from the coherence peaks or the
kink position to the superconducting valleys (arrows in
Fig.~\ref{fig4}(a)) are mainly 3$\;$meV for most of the spectra,
while some of the spectra show much larger gap sizes or even
two-gap features. In high-$T_\mathrm{c}$ superconductors,
sometimes a bosonic mode which exhibits as a peak feature at a
higher energy outside of the superconducting gap is found by STS
measurements \cite{29,30}. The second gap in Bi$_4$O$_4$S$_3$ resembles
the bosonic mode feature. However such high-energy peaks are quite
rare to occur in hundreds of measured spectra, so we just regard
it as a possible second gap. In order to know the average value of
the superconducting gap, we do the statistic analysis to the gap
size $\Delta$ taken from 400 spectra and present in
Fig.~\ref{fig4}(b). One can see that the mean gap value
$\overline{\Delta}$ is about 3$\;$meV. Considering the bulk
superconducting transition temperature
$T_\mathrm{c}^\mathrm{onset}$ of 4.2$\;$K, we get the ratio
$2\overline{\Delta}/k_\mathrm{B}T_\mathrm{c}\sim 16.6$, which is
almost 5 times of the value given by the BCS theory in the weak
coupling regime. It is even higher than the values of most
high-$T_\mathrm{c}$ superconductors. This suggests the very strong
coupling superconductivity in the superconductor. Since the
scattering is really strong in the polycrystalline sample, it is
very difficult to judge the pairing symmetry from the fitting to
the spectra. As shown in Fig.~\ref{fig4}(b), the gap size can
extend to a very large value, e.g., exceeding 10$\;$meV. Such
inhomogeneity of SC gap sizes needs to be verified by other
experiment tools which may make it an interesting new material.

\begin{figure}
\includegraphics[width=8cm]{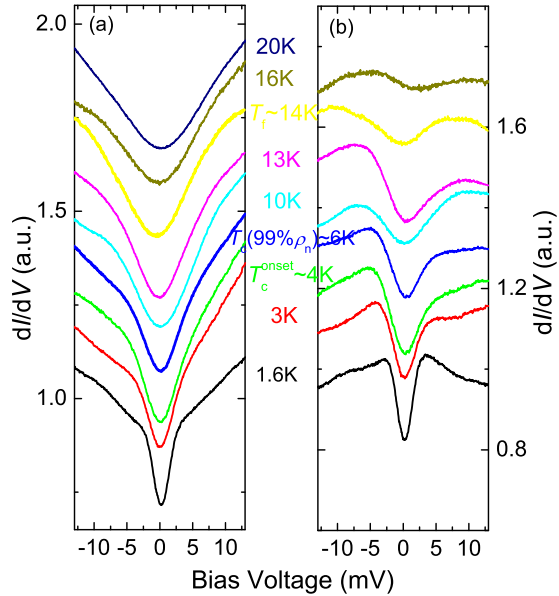}
\caption {(color online) (a) The evolution of the tunnelling
spectra taken at temperatures from 1.6 K to 20 K. The spectra are
displaced vertically for clarity. (b) The STS normalized by the
one measured in normal state (at 20 K). One can see that the
gapped feature vanishes at about 14$\;$K which is much higher than
the critical temperature for bulk superconductivity
($T_\mathrm{c}(99\%\rho_\mathrm{n})\sim 6\;$K), as shown by the
blue curve.} \label{fig5}
\end{figure}

In Fig.~\ref{fig5}(a), we show the temperature evolution of the
STS spectra obtained by warming up the samples from 1.6$\;$K
through $T_\mathrm{c}$ to 20$\;$K. One can see that the
superconducting feature marked by the depression of the density of
states near the Fermi energy persists at temperatures above the
bulk $T_\mathrm{c}(99\%\rho_\mathrm{n})\sim6\;$K. Such a feature
disappeared at temperature above some fluctuation temperature
$T_\mathrm{f}=14\;$K leaving only a V-shaped background. If we use
the spectrum measured at $20\;$K as the background, we can obtain
the normalized curves at different temperatures as shown in
Fig.~\ref{fig5}(b). Superconducting feature is weakened with the
increasing of temperature and evolve to a continuous background at
the temperature above $T_\mathrm{f}$. In traditional
superconductors, the superconducting gapped features on tunnelling
spectra vanish at the temperature just above $T_\mathrm{c}.$ \cite{31}
In contrast such feature in cuprates could exist at the
temperature far beyond $T_\mathrm{c}$, which has been regarded as
the pseudogap effect \cite{16}. Even in some iron pnictides, the gapped
feature was observed to extend to a very high temperature \cite{32} and
was explained by the presence of possible pseudogap effect \cite{33}.
The pseudogap effect can also be observed from the kink in
resistive curve in cuprates \cite{13}. Because we cannot find any trace
of pseudogap from the transport measurements, this effect at high
temperature is supposed to be the SC fluctuation instead of the
pseudogap. In addition, the estimation is consistent with excess
conductivity at the temperature above bulk $T_\mathrm{c}$. If
using $T_\mathrm{f}=14\;$K as the pairing temperature, we get the
ratio $2\overline{\Delta}/k_\mathrm{B}T_\mathrm{f}\sim5.0$ which
is still a large value but comparable with the value calculated
from the SC gap and the pseudogap temperature in cuprates \cite{3}. It
should be noted that some SC gap values could extend to very high,
i.e., larger than 7$\;$meV, which gives a much larger value of
$2\overline{\Delta}/k_\mathrm{B}T_\mathrm{f}$. The detailed reason
for this large energy gap remains unresolved.

\begin{figure}
\includegraphics[width=9cm]{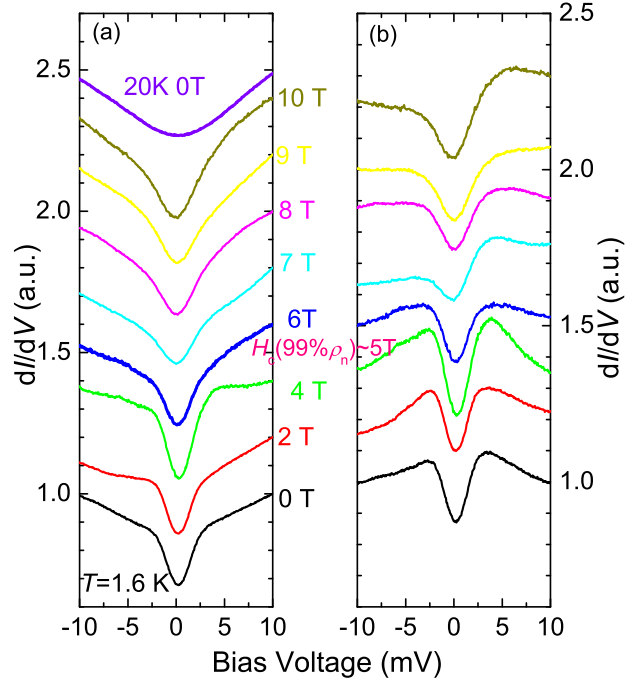}
\caption {(color online) (a) The evolution of the tunnelling
spectra at different magnetic fields up to 10$\;$T at 1.6$\;$K.
The spectra are displaced vertically for clarity. The spectrum
taken at 20$\;$K and 0$\;$T is also shown for comparison. (b) The
STS curves normalized by the one measured in normal state (at
20$\;$K and 0$\;$T). One can see the suppression of DOS remains at
fields above the bulk upper critical field
$H_\mathrm{c}(99\%\rho_\mathrm{n}) \sim5\;$T.} \label{fig6}
\end{figure}

Figure~\ref{fig6}(a) shows the STS spectra taken at different
magnetic fields at the same temperature $1.6\;$K. The bulk upper
critical field $\mu_0H_\mathrm{c2}$ judged from the $99\%$ of the
normal state resistance at $1.6\;$K is about $5\;$T. The
suppression to the DOS on the spectra are apparently without any
variation when crossing this bulk transition temperature, and it
is more clear for the normalized spectra by dividing out the
background spectrum taken at 20$\;$K and $0\;$T as shown in
Fig.~\ref{fig6}(b). As described above, the gapped feature on the
spectra existing above $T_\mathrm{c}(99\%\rho_\mathrm{n})$ is
consistent with the picture of fluctuating superconductivity.
Since the spectra at high magnetic fields are similar to those
taken at zero field but at high temperatures, this suppression of
DOS near Fermi energy observed above bulk
$H_\mathrm{c}(99\%\rho_\mathrm{n})$ can also be attributed to the
SC fluctuation and preformed Cooper pairs.

\begin{figure}
\includegraphics[width=8cm]{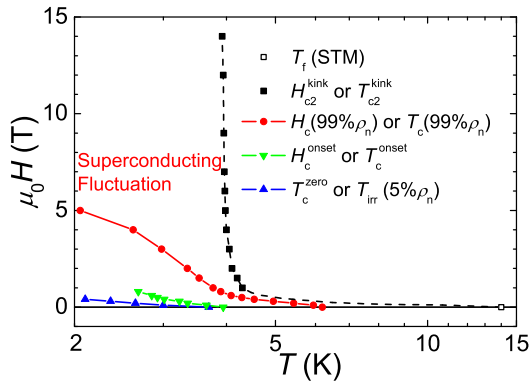}
\caption {(color online) Phase diagram derived from the resistive
transition curves and the STS data. Semi-log plot is used to make
the phase diagram more clear. The transition point judging from
the $99\%$ of the normal resistance gives rise to the upper
critical field $H_\mathrm{c2}$. The curve
$H_\mathrm{c2}^\mathrm{kink}$ shows the point determined from the
kink point of the resistivity versus temperature which denotes the
superconducting fluctuation property. Such fluctuation behavior
with excess conductivity is proved by the STS data and extends to
as high as 14 K at 0 T. The dashed line is a guide for eyes.}
\label{fig7}
\end{figure}

\section{Discussion}

Next we present a phase diagram based on the transport and STS
measurements in Fig.~\ref{fig7} and give discussions on the
possible mechanism of superconductivity. The SC transition point
of critical field $H_\mathrm{c}(99\%\rho_n)$ is shown by the red
filled circles. The bulk superconductivity is established in a
very small area covered by the irreversibility line
$T_\mathrm{irr}$ (blue up-triangles). The large area between them
indicates a strong SC phase fluctuation. This is actually
consistent with the theoretical expectation because the electronic
system has one dimensional feature($p_\mathrm{x}$ and
$p_\mathrm{y}$). The bulk superconductivity is established between
the one dimensional fluctuating superconducting chains. The curve
marked with $H_\mathrm{c}^\mathrm{onset}$ gives the upper critical
field determined using the usual crossing point of the normal
state background and the extrapolated line of the steep resistive
transition part. The most puzzling point is the kink appearing in
the $\rho$ vs. $T$ data at a high magnetic field. The curve marked
with $H_\mathrm{c2}^\mathrm{kink}$ shows the critical field
determined from the kinky point of the resistive data shown in
Fig.~\ref{fig3}(c), by following the trace of the arrowed red line
there. We add the fluctuation temperature $T_\mathrm{f}$ from the
tunnelling spectrum at zero magnetic field to the phase diagram,
and get a wide fluctuation region at zero magnetic field. Since
this line traces very well to the transition point marked by
$H_\mathrm{c}(99\%\rho_\mathrm{n})$ in the low field and high
temperature region, we naturally attribute it to the existence of
residual Cooper pairs. If this kink can be interpreted as the
onset for the pairing, that would indicate a very strong pairing
strength or gap, which can be inversely supported by the tunneling
data. In a simple BCS picture, we have
$H_\mathrm{c2}=(\pi\Phi_0/2\hbar^{2}v_\mathrm{F}^{2})\Delta^{2}$,
where $\Phi_0$ is the flux quanta, $v_\mathrm{F}$ is the Fermi
velocity. Such a strong pairing needs certainly a reasonable
cause, which exceeds the limit of the simple phonon mediated
pairing picture. By taking account of the weak correlation effect
in the Bi 6p electrons, some other novel mechanism, such as the
valence fluctuation of the Bi$^{2+}$ and Bi$^{3+}$, may play an
important role in this new superconductor.

\section{Conclusions}

In summary, we perform the resistive and scanning tunnelling
spectroscopy measurements on the new BiS$_2$ based superconductor
Bi$_4$O$_4$S$_3$. A weak insulating behavior is induced in the
normal state when a high magnetic field is applied. This can be
induced either by an adjacent competing order, or the very shallow
$p_x$ and $p_y$ band and small Fermi energy. A kink appears on the
temperature dependence of resistivity at all high magnetic fields
when the bulk superconductivity is completely suppressed. This
kink can be regarded as the presence of local pairing, or the
upper critical field $H_\mathrm{c2}$(T). The SC fluctuation region
from the STS measurement extends to about 14 K although the bulk
superconducting transition temperature is only about 3.7 K. The
gapped feature near the Fermi energy can also extend to a high
magnetic field (~ 5 T), which is consistent with the resistive
measurements, again indicating a strong superconducting
fluctuation. From the tunnelling spectra, a mean superconducting
gap of 3$\;$meV is widely observed, which leads to a very high
ratio of 2$\Delta/k_\mathrm{B}T_\mathrm{c}\approx 16.6$,
suggesting strong coupling superconductivity.

We appreciate the useful discussions with WANG Fa, FU Liang, WANG
QiangHua and LI JianXin. This work was supported by the 973
Project of the Ministry of Science and Technology of China (Grant
Nos. 2011CBA001002, 2010CB923002, and 2012CB821403), the National
Natural Science Foundation of China (Grant No. 11034011), the
Program for New Century Excellent Talents in University (Grant No.
NCET-12-0255), and A Project Funded by the Priority Academic
Program Development of Jiangsu Higher Education Institutions.

$^{\dag}$huanyang@nju.edu.cn, $^*$hhwen@nju.edu.cn


\begin{thebibliography}{99}
\bibitem{1}Bednorz J G and Muller K A. Possible high T$_c$
superconductivity in the Ba-La-Cu-O system. Zeitschrift fur Physik
B Condensed Matter 1986, 64: 189-193

\bibitem{2} Kamihara Y, Watanabe T, Hirano M, et al. Iron-Based
Layered Superconductor La[O$_{1-x}$F$_x$]FeAs (x=0.05-0.12) with
T$_c$=26 K. J. AM. CHEM. SOC 2008, 130: 3296-3297

\bibitem{3} Fischer {\O}, Kugler M, Maggio-Aprile I, et al. Scanning
tunneling spectroscopy of high-temperature superconductors. Rev.
Mod. Phys, 2007, 79: 353-419


\bibitem{4} Hoffman J E. Spectroscopic scanning tunneling microscopy
insights into Fe-based superconductors. Rep. Prog. Phys, 2011, 74:
124513

\bibitem{5} Wolf E, Principles of Electron Tunneling Spectroscopy,
Oxford University Press, New York, 1985

\bibitem{6} Xu Z A, Ong N P, Wang Y, et al. Vortex-like excitations
and the onset of superconducting phase fluctuation in underdoped
La$_{2-x}$Sr$_x$CuO4. Nature(London), 2000, 406: 486-488

\bibitem{7} Wang Y Y, Li L, Ong N P. Nernst effect in high-T$_c$
superconductors. Phys. Rev. B, 2006, 73: 024510

\bibitem{8} Wang Y Y, Li L, Naughton M J, et al. Field-Enhanced
Diamagnetism in the Pseudogap State of the Cuprate
Bi$_2$Sr$_2$CaCu$_2$O$_{8+\delta}$ Superconductor in an Intense
Magnetic Field. Phys. Rev. Lett, 2005, 95: 247002

\bibitem{9} Wen H H, Mu G, Luo H Q, et al. Specific-Heat Measurement
of a Residual Superconducting State in the Normal State of
Underdoped Bi$_2$Sr$_{2-x}$La$_x$CuO$_{6+\delta}$ Cuprate
Superconductors. Phys. Rev. Lett, 2009, 103: 067002

\bibitem{10} Rourke P M C, Mouzopoulou I, Xu X F, et al.
Phase-fluctuating superconductivity in overdoped
La$_{2-x}$Sr$_x$CuO$_4$, Nature Physics, 2011, 7: 455-458

\bibitem{11} Zhu Z W, Xu Z A, Lin X, et al. Nernst effect of a new
iron-based superconductor LaO$_{1-x}$F$_x$FeAs. New J. Phys, 2008,
10: 063021

\bibitem{12} Pourret A, Malone Liam, Antunes A B, et al. Strong
correlation and low carrier density in
Fe$_{1+y}$Te$_{0.6}$Se$_{0.4}$ as seen from its thermoelectric
response. Phys. Rev. B, 2011, 83: 020504

\bibitem{13} Timusk T, Statt B. The pseudogap in high-temperature
superconductors: an experimental survey. Rep. Prog. Phys, 1999,
62: 61-122

\bibitem{14} Kondo T, Hamaya Y, Palczewski A D, et al. Disentangling
Cooper-pair formation above the transition temperature from the
pseudogap state in the cuprates. Nature Physics, 2011, 7: 21-25

\bibitem{15} Grbi\'{c} M S, Po\v{z}ek M, Paar D, et al. Temperature range of
superconducting fluctuations above T$_c$ in
YBa$_2$Cu$_3$O$_{7-\delta}$ single crystals. Phys. Rev. B, 2011,
83: 144508

\bibitem{16} Renner C, Revaz M, Genoud J Y, et al. Pseudogap
Precursor of the Superconducting Gap in Under- and Overdoped
Bi$_2$Sr$_2$CaCu$_2$O$_{8+\delta}$. Phys. Rev. Lett, 1998, 80:
149-152

\bibitem{17} Anderson P W, Lee P A, Randeria M, et al. The physics
behind high-temperature superconducting cuprates: the 'plain
vanilla' version of RVB. J. Phys: Condens. Matter, 2006, 16:
R755-R769

\bibitem{18} Scalapino D J. The case for d$_{x^2-y^2}$ pairing in the
cuprate superconductors. Physics Report, 1995, 250:329-365. Moriya
T, Ueda U. Antiferromagnetic spin fluctuation and
superconductivity. Rep. Prog. Phys, 2003, 66: 1299-1341. Monthoux
P, Pines D, Lonzarich G G. Superconductivity without phonons,
Nature (London) 2007, 450: 1177-1183

\bibitem{19} Ni N, Tillman M E, Yan J Q, et al. Effects of Co
substitution on thermodynamic and transport properties and
anisotropic H$_{c2}$ in Ba(Fe$_{1-x}$Co$_x$)$_2$As$_2$ single
crystals. Phys. Rev. B, 2008, 78: 214515

\bibitem{20} Si Q M, Steglich F. Heavy Fermions and Quantum Phase
Transitions. Science, 2010, 329: 1161-1166

\bibitem{21} Dressel M. Quantum criticality in organic conductors?
Fermi liquid versus non-Fermi-liquid behaviour. J. Phys. Condens.
Matter, 2011, 23: 293201

\bibitem{22} Mizuguchi Y, Fujihisa H, Gotoh Y, et al. BiS$_2$-based
layered superconductor Bi$_4$O$_4$S$_3$. Phys. Rev. B, 2012, 86:
220510

\bibitem{23} Mizuguchi Y, Demura S, Deguchi K, et al.
Superconductivity in Novel BiS$_2$-based superconductor
LaO$_{1-x}$F$_x$BiS$_2$. J. Phys. Soc. Jpn, 2012, 81: 114725

\bibitem{24} Demura S, Mizuguchi Y, Deguchi K, et al. New member of
BiS$_2$-based superconductor NdO$_{1-x}$F$_x$BiS$_2$. J. Phys.
Soc. Jpn, 2013, 82: 033708

\bibitem{25} Xing J, Li S, Ding X X, et al. Superconductivity appears
in the vicinity of semiconducting-like behaviour in
CeO$_{1-x}$F$_x$BiS$_2$. Phys. Rev. B, 2012, 86: 214518

\bibitem{26} Usui H, Suzuki K, Kuroki K. Minimal electronic models
for superconducting BiS$_2$ layers. Phys. Rev. B, 2012, 86:
220501(R)

\bibitem{27} Chen B X and Uher. Transport properties of Bi$_2$S$_3$
and the ternary bismuth sulfides KBi$_{6.33}$S$_{10}$ and
K$_2$Bi$_8$S$_{13}$. Chem Matter, 1997, 9: 1655-1658

\bibitem{28} Clogston A M. Upper Limit for the Critical Field in Hard
Superconductors. Phys. Rev. Lett, 1962, 9: 266-267

\bibitem{29} Niestemski F C, Kunwar S, Zhou S. A distinct bosonic
mode in an electron-doped high-transition-temperature
superconductor. Nature, 450: 1058-1061

\bibitem{30} Wang Z Y, Yang H, Fang D L, et al. Close relationship
between superconductivity and the bosonic mode in
Ba$_{0.6}$K$_{0.4}$Fe$_2$As$_2$ and
Na(Fe$_{0.975}$Co$_{0.025}$)As. Nature Physics 2013, 9: 42-48

\bibitem{31} Pan S H, Hudson E W and Davis J C. Vacuum tunneling of
superconducting quasiparticles from atomically sharp scanning
tunneling microscope tips. Appl. Phys. Lett, 1998, 73: 2992

\bibitem{32} Yang H, Wang Z Y, Fang D L, et al. Unexpected weak
spatial variation in the local density of states induced by
individual Co impurity atoms in superconducting
Na(Fe$_{1-x}$Co$_x$)As crystals revealed by scanning tunneling
spectroscopy. Phys. Rev. B, 2012, 86: 214512

\bibitem{33} Zhou X D, Cai P, Wang A F, et al. Evolution from
Unconventional Spin Density Wave to Superconductivity and a
Pseudogaplike Phase in NaFe$_{1-x}$Co$_x$As. Phys. Rev. Lett,
2012, 109: 037002


\end{thebibliography}
\end{document}